\documentclass[aps,prl,twocolumn,showpacs]{revtex4} 
\usepackage{graphics,epsfig}
\newcommand{\be}{\begin{equation}}
\newcommand{\ee}{\end{equation}}
\newcommand{\ba}{\begin{eqnarray}}
\newcommand{\ea}{\end{eqnarray}}
\newcommand{\lb}{\label}

\newcommand{\half}{\frac{1}{2}}

\newcommand{\nn}{\nonumber}
\newcommand{\bwt}{ \begin{widetext}}
\newcommand{\ewt}{ \end{widetext}}
\begin{document}
\title{ A Manifestly Noncovariant Theory of Gravitation}
\author{ Patricio.S. Letelier }

\email{e-mail: letelier@ime.unicamp.br}
 \affiliation{
 Departamento de Matem\'atica Aplicada-IMECC,
Universidade Estadual de Campinas,
13083-859 Campinas,  S.P., Brazil}
\begin{abstract}
We study a noncovariant theory of gravitation based on the Lagrangian density  $\sqrt{-g})^\omega R$, where $\omega$ is a constant. In particular, we study solutions  that for $\omega=1$ reduces to the de Sitter,  Kasner and  LFRW (with perfect fluid with as a source). We also consider spherically symmetric solutions.
\end{abstract}
\maketitle
\setlength{\parindent}{3em}
\section{introduction}
Recently there has being a lot of attention to theories of gravitation that break the equivalence between the space
and the time. Examples of these theories are: The TeVeS  (Tensor Vector Scalar) \cite{teves} theory that is a covariant version  of MOND (Modificated Newtonian Dynamics) \cite{ mond} that is thought
as  a modification of Newtonian  gravitation in order to solve the problem of the galaxy rotation curves without the introduction of dark matter. In TeVeS a fundamental timelike vector field  is introduced that gives a priviledged time direction.  The   Einstein aether
theory \cite{aether1} where a  dynamical  unit timelike vector field is introduced to  breaks local Lorentz symmetry.  The main motivation for this theory is  the suspection  that the vacuum in quantum gravity may determine a preferred
rest frame at the microscopic level.  The vector field defines a congruence of timelike curves filling all of spacetime, like an omnipresent fluid, and so has been named  ``aether". And the  theory of gravity
proposed by Hor\v ava  \cite{horava}, inspired by condensed matter
models of dynamical critical systems. It has manifest
three-dimensional spatial general covariance and time reparametrization
invariance. It is described in the language of the   Arnowitt-Deser-Misner (ADM)  canonical Hamiltonian  formulation
of general relativity, but in which Einstein
gravity is modified adding a new constant  so that the full underlying four dimensional
covariance is broken. But,  it is restored   at  the  infrared large distance
limit, i.e., for a particular value of the above mentioned constant.  The motivation for this theory is the  construction of a  renormalizable theory in
3+1 dimensions.   The relation between these last two theories is studied in \cite{aether2}. 

The purpose of this letter is to study t a new noncovariant modification of 
General Relativity (GR) that contains a new constant such that when the constant takes a particular value we 
recover  the usual GR. Our starting point is the action for the gravity part of the theory,
\be
 S_G=  \int (\sqrt{-g})^\omega(R+2\Lambda)d^4 x,     \lb{ag}
\ee
where $\Lambda$  the cosmological constant  and $\omega$ is a new constant.  We use geometrical units $8\pi G=c=1$ and metric signature $(+ - - -)$. The previous action is  invariant under a general  change of coordinates only    when $\omega=1$. Therefore, for  $\omega\not= 1$ the variation of this action yield noncovariant field equations  equations. For the matter we choose the action,
\be
S_M=  \int (\sqrt{-g})^\sigma L_Md^4 x,  \lb{am}
\ee
where $L_M$  is the usual  Lagrangian  that describes the matter content  of the universe, and $\sigma$ is a constant, we shall consider the two natural possibilities, that either $\sigma$ takes the value  $1$ or   $\omega$. For the first value will  have the usual coupling of matter used  in GR and the second value is a natural choice compatible with  the introduction of the parameter that breaks the  covariance  in (\ref{ag}) . From the functional derivative, $\delta(S_G+2S_M)/\delta g^{\mu\nu}=0,$ we find
\begin{widetext}
\be
R_{\mu\nu}-\frac{\omega}{2}R+\omega\Lambda g_{\mu\nu}+(\omega-1)K_{\mu\nu}= -(\sqrt{-g})^{\sigma-\omega}T_{\mu\nu} -(1-\sigma)   (\sqrt{-g})^{\sigma-\omega}  g_{\mu\nu}L_M,  \lb{fe}
\ee
\end{widetext}
where the object $K_{\mu\nu}$ is constructed with $g_{\mu\nu}$ and its first derivatives. We note that 
 $(\sqrt{-g})^\omega R$ is linear in the second derivatives of the metric. So  these second derivatives 
can be eliminated by  the usual integration by parts. The explicit form of the  object $K_{\mu\nu}$, that is not a tensor,  is quite large and cumbersome and will be presented elsewhere. The  tensor $T_{\mu\nu}$ is  the usual metric energy-momentum tensor [$\sqrt{-g}T_{\mu\nu}=2\delta (\sqrt{-g}L_M)/\delta g^{\mu\nu}$].  From eq. (\ref{fe}) we see that  when  $\omega=1$  we recover the usual GR  ($ \sigma$ takes the values either  one or $\omega$).  So we will pay special attention to the transition from $\omega\not= 1$, early universe,  to  $\omega=1$,  present era.

In this letter we shall consider three solutions with cosmological interest,  a de Sitter like solution, a Kasner like solution, and a Lema\^itre-Friedman-Roberson-Walker  (LFRW) like solution with a $p=\gamma \rho$ fluid as a source as well as some spherically symmetric solutions.

For the LFRW metric with flat spatial sections,
\be
ds^2=dt^2- a^2(t)(dx^2+dy^2+dz^2),   \lb{lfrw}
\ee
we find that the variation of $S_G$ gives us,
\be
2\left(3\omega-2\right)\frac{d ^2a}{d t^2}a+\left(3\omega-2\right)^2\left(\frac{d a}{d t}\right)^2-\Lambda a^2\omega=0. \lb{dseq}
\ee
Looking for a solution of the form $ a=e^{\alpha t} $, where $\alpha$ is a constant we get
\be
\alpha=\sqrt{\frac{\Lambda}{3( 3\omega-2)}}.
\ee
The symmetry breaking parameter $\omega$ introduces a renormalization of the cosmological constant.
Suppose that in the early universe $\omega$ is  close to $2$, in this  case we can have an effective cosmological constant as large as wished. The for $\omega=1$ we recover the usual expansion rate. This transition can give
us a hint to solve the problem of the size of the cosmological constant computed in  quantum field theory. 

The Now we shall consider a Bianchi type I metric,
\be
ds^2=N^2(t)dt^2- a^2(t)dx^2- b^2(t) dy^2- c^2(t) dz^2. \lb{bI}
\ee
In the study of the generic  cosmological singularity of the solution of the Einstein equations this metric with $N=1$  plays an essential role \cite{bkl}.

From $\delta S_G=0$ we get 
\bwt
\ba
&&\frac{d a}{d t}\frac{d b}{d t}c+\frac{d a}{d t}\frac{d c}{d t}b+\frac{d b}{d t}\frac{d c}{d t}a+\nn\\
&&
\left(\frac{d ^2a}{d t^2}ab^2c^2+ \left( \frac{d a}{d t}\right)^2b^2c^2+3\frac{d a}{d t}\frac{d b}{d t}abc^2+3\frac{d a}{d t}\frac{d c}{d t}ab^2c+\frac{d ^2b}{d t^2}a^2bc^2+\left(\frac{d b}{d t}\right)^2a^2c^2+   \right. \nn\\
&& \left. 3\frac{d b}{d t}\frac{d c}{d t}a^2bc+\frac{d ^2c}{d t^2}a^2b^2c+\left(\frac{d c}{d t}\right)^2a^2b^2\right) \frac{ \left(\omega-1\right)}{abc}=0 ,\lb{vn}
\ea
\ba
&& Na^2bc\left(\frac{d N}{d t}\frac{d b}{d t}c+\frac{d N}{d t}\frac{d c}{d t}b-\frac{d ^2b}{d t^2}Nc-\frac{d b}{d t}\frac{d c}{d t}N-\frac{d ^2c}{d t^2}Nb\right)+ \nn\\
&& bc
 \left(-\frac{d ^2N}{d t^2}Na^2bc+2\left(\frac{d N}{d t}\right)^2a^2bc+2\frac{d N}{d t}\frac{d a}{d t}Nabc+\frac{d N}{d t}\frac{d b}{d t}Na^2c+\frac{d N}{d t}\frac{d c}{d t}Na^2b-2\frac{d ^2a}{d t^2}N^2abc+\right. \nn\\
&& \left. \left(\frac{d a}{d t}\right)^2N^2bc-2\frac{d a}{d t}\frac{d b}{d t}N^2ac-2\frac{d a}{d t}\frac{d c}{d t}N^2ab-2\frac{d ^2b}{d t^2}N^2a^2c-3\frac{d b}{d t}\frac{d c}{d t}N^2a^2-2\frac{d ^2c}{d t^2}N^2a^2b\right)\left(\omega-1\right)+ \nn\\
&& \left(-\left(\frac{d N}{d t}\right)^2a^2b^2c^2-2\frac{d N}{d t}\frac{d a}{d t}Nab^2c^2-2\frac{d N}{d t}\frac{d b}{d t}Na^2bc^2-2\frac{d N}{d t}\frac{d c}{d t}Na^2b^2c-\left(\frac{d a}{d t}\right)^2N^2b^2c^2-  \right.  \nn\\
&& \left.  2\frac{d a}{d t}\frac{d b}{d t}N^2abc^2-2\frac{d a}{d t}\frac{d c}{d t}N^2ab^2c- \left(\frac{d b}{d t}\right)^2N^2a^2c^2-2\frac{d b}{d t}\frac{d c}{d t}N^2a^2bc-\left(\frac{d c}{d t}\right)^2N^2a^2b^2\right)\left(\omega-1\right)^2=0,
\lb{va}\\
&&
Do  \;\;\;     a \rightarrow b \rightarrow c\rightarrow a   \;\;\; in  \;\;\;  (\ref{va}) ,\lb{vb}\\
&&
Do  \;\;\;     a \rightarrow b \rightarrow c\rightarrow a   \;\;\;  in   \;\;\;  (\ref{vb}). \lb{vc}
\ea
\ewt
Now choosing the gauge $N=1$ we have that these last three equations reduce to
\bwt
\ba
 &&a^2bc\left(\frac{d ^2b}{d t^2}c+\frac{d b}{d t}\frac{d c}{d t}+\frac{d ^2c}{d t^2}b \right) +\nn\\
&&bc\left(2\frac{d ^2a}{d t^2}abc-\left(\frac{d a}{d t}\right)^2bc+2\frac{d a}{d t}\frac{d b}{d t}ac+2\frac{d a}{d t}\frac{d c}{d t}ab+2\frac{d ^2b}{d t^2}a^2c+3\frac{d b}{d t}\frac{d c}{d t}a^2+2\frac{d ^2c}{d t^2}a^2b\right)\left(\omega-1\right)+ \nn \\
&&\left(\left(\frac{d a}{d t}\right)^2b^2c^2+2\frac{d a}{d t}\frac{d b}{d t}abc^2+2\frac{d a}{d t}\frac{d c}{d t}ab^2c+\left(\frac{d b}{d t}\right)^2a^2c^2+2\frac{d b}{d t}\frac{d c}{d t}a^2bc+\left(\frac{d c}{d t}\right)^2a^2b^2\right)\left(\omega-1\right)^2=0  \lb{van1}, \\
&&Do  \;\;\;     a \rightarrow b \rightarrow c\rightarrow a   \;\;\; in  \;\;\;  (\ref{van1}), \lb{vbn1}\\
&&Do  \;\;\;     a \rightarrow b \rightarrow c\rightarrow a   \;\;\;  in   \;\;\;  (\ref{vbn1}) .\lb{vcn1}
\ea
\ewt
Now looking for a solution of the form $a=t^{p_1}, b=t^{p_2}, c=t^{p_3}$ we find 
\bwt
\ba
&&p_1 p_2+p_2 p_3+p_1 p_3+(\omega-1)(2p^{2}_1+2p^{2}_2+2p^{2}_3
-p_1-p_2-p_3 +3p_1 p_2+3p_2 p_3+3p_3 p_1)=0,  \lb{k1}\\
&&[\omega(p_1+p_2+p_3)-1](p_i-p_j)=0.   \;\;\; (i\not =j=1,2,3) \lb{k2}
\ea
\ewt
Thus, the constants $p_i$ satisfy,
\be
\omega(p_1+p_2+p_3)=1, \;\;\  p^{2}_1+p^{2}_2+p^{2}_3=(2\omega-1)/\omega^2.  \lb{knc}
\ee
Solving the above equations in terms of $p_1$, we find
 \ba
&&2\omega p_2=1-\omega p_1 +\sqrt{\Delta}, \\
&& 2\omega p_3=1-\omega p_1 -\sqrt{\Delta}, \lb{eqp}\\
&& \Delta\equiv -3[\omega P_1-(1-2\sqrt{3\omega-2}/3)]\times \nn \\
&& \;\;\;\;\;\;\;\;\; [\omega P_1-(1+2\sqrt{3\omega-2}/3)].
\ea
Thus to have a real solution we need $\omega \geq 2/3$ and 
\be 
3\omega p_1\geq1-2\sqrt{3\omega -2}, \;\;\;\;   3\omega p_1\leq 1+2\sqrt{3\omega -2}.\lb{ineq}
\ee
We find   for $\omega>1$ a similar behavior than in GR,  stretching and shrinking of the space in different directions governed by the signs of the exponents $p_i$. But,  for
$ 2/3<\omega<1$ we find that we can have that all the exponents positive, i.e., expansion in all directions, e.g.,  for $\omega \in[0.7,0.8] $ and $p_1\in[0.2,0.3]$ we 
have $p_i >0$ ($i=1,2,3$). This suggest that  near the generic singularity  for $\omega \not=1$ the spacetime will differ
 from the one of GR studied in \cite{bkl}.

Now we shall consider a LFRW universe filled with an irrotational  perfect fluid with $p=\gamma \rho$ equation of state. The Lagrangian for this fluid can be expressed in terms of the velocity potential, $\Phi$,  as  \cite{tt}
$L_M=p=\half( \partial^{\alpha}\Phi\partial_\alpha  \Phi)^{(1+\gamma)/2\gamma}$. The four-velocity takes de form $u_\mu=\partial_\mu\Phi/\sqrt{ \partial^{\alpha}\Phi\partial_\alpha  \Phi }$. The variation of $S_G+2S_M$ gives as
\ba
&&a^{3\sigma}\left(\frac{d \Phi}{dt}\right) ^{\frac{1}{\gamma}}=K\lb{phi1},\\
&& 4\left(3\omega-2\right)a^{3\omega+1}\frac{d ^2a}{d t^2}+
 2\left(3\omega-2\right)^2a^{3\omega}\left(\frac{d a}{d t}\right)^2  \nn\\
&&\;\;\;\;\;\;\;\ +\sigma a^{3\sigma+2}
  \left(\frac{d \Phi}{dt}\right)  ^{\frac{1+\gamma}{\gamma}}=0, \lb{af}
\ea
where $K$ is an integration constant and $\Phi$ is a function of $t$ only.
Form the equations above and the ansatz $a=t^\alpha$ we find
\be
2\alpha  \left(3\alpha\omega-2\right)\left(3\omega-2\right) t^{3\left(\gamma\sigma+\omega\right)\alpha}+\sigma K^{1+\gamma}t^2=0.\ee
Thus  
\be
 \alpha=\frac{2}{3\gamma\sigma+3\omega}, \;\;
 K^{1+\gamma}=\frac{8\left(3\omega-2\right)\gamma}{3\left(\gamma\sigma+\omega\right)^2}.\lb{alk}
\ee
For the density and the Hubble ``constant"   ($ H\equiv \frac{1}{a}\frac{da}{dt}$)  we get, respectively,
\be 
\rho=\frac{4(3\omega-2)}{3(\gamma\sigma+\omega)^2} t^{
-\frac{2\sigma(1+\gamma)}{\gamma\sigma +\omega}}, \;\;
H=\frac{2}{3(\gamma\sigma+\omega)t}. \lb{rhoh}
\ee
Therefore $\omega>2/3$ to have a positive density. For this limit value of $ \omega$ and $\sigma=\omega$ we have a expansion that is  $\%50$ greater that the predicted in GR. For $\omega>1$ we have an expansion that is slower than in GR.

Now we shall study some spherically symmetric solutions of eq. (\ref{fe}),
We shall consider the metric,
\be
ds^2=e^{A(r)}dt^2 -e^{B(r)}dr^2-U^2(r)(d\vartheta^2+\sin^2(\vartheta)d\varphi^2). \lb{sph}
\ee
For  $A=B=0$ the evolution equations reduce to
\be
\left(4\omega-3\right)\left(U \frac{d ^2U}{d r^2}+ \left(\omega-1\right)\left(\frac{d U}{d r}\right)^2\right)+ 1-\omega=0. \lb{streq}
\ee
A solution to this equations is    $U=r/\sqrt{4\omega-3}$. We have found other exact solutions  that we shall presented elsewhere. The Einstein tensor for this spacetime  reduces to $G^{t}_{t}=G^{r}_{r}=4(\omega-1)/r^2$. This spacetime in GR has already appeared and was related to a cloud of cosmic strings  \cite{str} and to monopoles \cite{mo}.

Now we consider the case $U=r$. The variation of the action (\ref{ag}) yield,
\bwt
\ba
&& 4\left(-\frac{d B}{d r}r-e^B+1\right)+\left(4\frac{d ^2A}{d r^2}r^2+\left(\frac{d A}{d r}\right)^2r^2-2\frac{d A}{d r}\frac{d B}{d r}r^2+8\frac{d A}{d r}r+2\frac{d ^2B}{d r^2}r^2-\left(\frac{d B}{d r}\right)^2r^2-4\frac{d B}{d r}r-4e^B \right.   \nn\\
&&\left. +\mathrm{12}\right)\left(\omega-1\right)+ \left(\left(\frac{d A}{d r}\right)^2r^2+2\frac{d A}{d r}\frac{d B}{d r}r^2+8\frac{d A}{d r}r+\left(\frac{d B}{d r}\right)^2r^2+8\frac{d B}{d r}r+\mathrm{16}\right)\left(\omega-1\right)^2=0, \nn \\
 && 2\left(\frac{d A}{d r}r-e^B+1\right)+\left(\frac{d ^2A}{d r^2}r^2+\left(\frac{d A}{d r}\right)^2r^2+6\frac{d A}{d r}r-2e^B+\mathrm{10}\right)\left(\omega-1\right)=0.  \lb{sph2}
\ea
\ewt

\begin{figure}
\includegraphics[scale=0.32 , angle=-90]{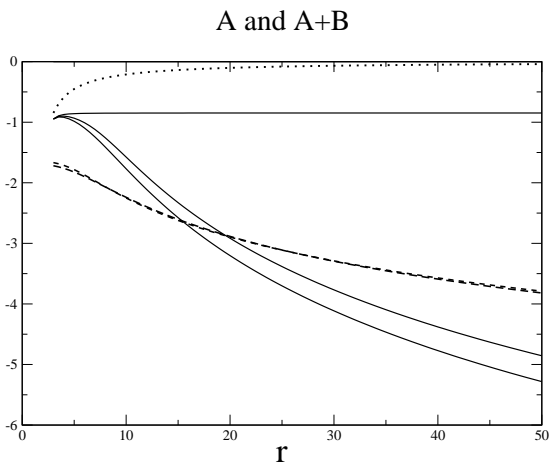}
\caption{The doted line is the graph of  $A(r)$  for $\omega=1$, $A(r)+B(r)=0$ in this case (Scharzschild solution). The second line is A for $\omega=1.5$, also $A+B=0$ in this case. The next two curves represent A for $\omega=1.6$ and $1.7$, respectively.  The segmented lines are $(A+B)/5$ for the previous two values of $ \omega$, repectively.}
\end{figure}

First we note that doing $\omega =1$ in eq. (\ref{sph2}) we recover the two first order differential equations that have as solutions $
e^A=e^{-B}=1-2m/r$,  i.e., the Schwarzschild  solution. Since the second order derivatives disappear in this case we have that the Schwazschild solution is a singular solution of  (\ref{sph2}). Using a fourth order Runge-Kutta algorithm we numerically solve the system (\ref{sph2}) with the  initial conditions $A(3)=-B(3)=\ln(3)$ and 
$dA(3)/dr=-dB(3)/dr=2/3$. These conditions  are obtained from  $A$ and $B$ at $r=3$ for the Schwarzschild solution with mass equal to one.   In Fig. 1 we present the solution of the above mentioned system for different values of  $\omega$. The doted line is the graph of  A(r)  for $\omega=1$,    i.e, the Schwarzshild solution;   $A(r)+B(r)=0$ in this case. The second line is  A for $\omega=1.5$, also $A+B=0$ in this case. We
note a very different behavior that in the precedent case, after $r=10$ $A$ is almost constant, we have a variation in the third decimal case only. The next two curves  are  A for $\omega=1.6$ and $1.7$,  they present a quite different behavior that the previous  cases, first they are decreasing functions of $r$, and $A\not=-B$.  The  graph of the function  $(A+B)/5$  for $\omega=1.6$ and $\omega=1.7$  are the two segmented  lines,  respectively.  We have that after the first step of integration the value of $B$ is   quite different from its initial value (not shown in the figure)  and the value of $A$ is similar to its initial value. This system of equations will be further explored  in another opportunity. 

In summary, the theory based in the action (\ref{ag}) present some interesting features like the change of the rate of expansion for cosmological models. And it is simple enough to present analytical solutions for the most important cosmological models.

\acknowledgments
The author thanks CNPq   (grant: 300327/2008-0) and FAPESP (grant:  2009/54572-0)    for partial financial support.

\end{document}